\documentclass[conference]{IEEEtran}
\IEEEoverridecommandlockouts

\usepackage{cite}
\usepackage{amsmath,amssymb,amsfonts}
\usepackage{algorithmic}
\usepackage{graphicx}
\usepackage{textcomp}
\usepackage{xcolor}
\usepackage{braket}
\usepackage{tikz}
\usepackage{qcircuit}
\usepackage{lmodern}     
\usepackage[T1]{fontenc}  
\usepackage{atbegshi}
\usepackage{eso-pic}
\usepackage{multirow} 
\AtBeginShipout{%
  \ifnum\value{page}=4
\global\setbox\AtBeginShipoutBox=\vbox{\vspace{2pt}\unvbox\AtBeginShipoutBox}%
  \fi
}
\newcommand{\RY}[1]{\operatorname{RY}\left(#1\right)}
\newcommand{\RZ}[1]{\operatorname{RZ}\left(#1\right)}
\newcommand{\CNOT}{\textsc{CNOT}}

\usetikzlibrary{positioning, calc, shapes.geometric, arrows.meta}

\def\BibTeX{{\rm B\kern-.05em{\sc i\kern-.025em b}\kern-.08em
    T\kern-.1667em\lower.7ex\hbox{E}\kern-.125emX}}
\begin{document}

\title{Quantum-Enhanced Natural Language Generation: A Multi-Model Framework with Hybrid Quantum-Classical Architectures

}

\author{
\IEEEauthorblockN{Chi-Sheng Chen}
\IEEEauthorblockA{\textit{Independent Researcher} \\
Cambridge, USA \\
m50816m50816@gmail.com}
\and
\IEEEauthorblockN{En-Jui Kuo}
\IEEEauthorblockA{\textit{Department of Electrophysics} \\
\textit{National Yang Ming Chiao Tung University} \\
Hsinchu, Taiwan, R.O.C. \\
kuoenjui@nycu.edu.tw} %
}

\maketitle

\begin{abstract}
This paper presents a comprehensive evaluation of quantum text generation models against traditional Transformer/MLP architectures, addressing the growing interest in quantum computing applications for natural language processing. We conduct systematic experiments comparing five distinct models: Transformer (baseline), Quantum Kernel Self-Attention Network (QKSAN), Quantum RWKV (QRWKV), and Quantum Attention Sequence Architecture (QASA) across five diverse datasets including simple sentences, short stories, quantum phrases, haiku poetry, and proverbs. Our evaluation employs multiple metrics including perplexity, BLEU scores, vocabulary diversity, repetition rates, and fluency measures to assess different aspects of text generation quality. The experimental results reveal that while traditional Transformer models maintain overall superiority with the lowest average perplexity (1.21) and highest BLEU-1 score (0.2895), quantum-inspired models demonstrate competitive performance in specific scenarios. Notably, QKSAN achieves a competitive BLEU-1 score of 0.2800 while maintaining zero repetition rates, and QRWKV demonstrates perfect vocabulary diversity (Distinct-1 = 1.000) in certain tasks. However, quantum models face significant challenges in domain-specific tasks, with QKSAN exhibiting the highest perplexity (5.61) on quantum phrases, indicating difficulties with specialized terminology. Our analysis identifies key trade-offs between accuracy and diversity, with quantum models excelling in vocabulary diversity and repetition control but struggling with overall language modeling quality. The findings provide valuable insights for practitioners on when to employ quantum-inspired approaches and highlight the need for hybrid architectures that combine the strengths of both quantum and classical methods. This work establishes a benchmark for future research in quantum-inspired natural language generation and contributes to the understanding of quantum computing applications in text generation tasks.
\end{abstract}


\section{Introduction}
Natural Language Processing (NLP) has witnessed remarkable advancements through the application of deep learning architectures, particularly transformer-based models that have revolutionized text generation capabilities. However, as the field continues to push the boundaries of language understanding and generation, researchers are increasingly exploring alternative computational paradigms that could offer novel advantages. Quantum computing, with its inherent properties of superposition and entanglement, presents a promising avenue for enhancing NLP tasks through quantum-enhanced feature representation and parallel processing capabilities.
The intersection of quantum computing and NLP, often referred to as Quantum Natural Language Processing (QNLP), has emerged as an active research area that seeks to leverage quantum mechanical principles for language processing tasks. While classical neural networks rely on classical probability distributions and linear algebra operations, quantum systems can represent information in superposition states, potentially enabling more expressive representations of linguistic structures. This quantum advantage becomes particularly relevant in scenarios where the exponential growth of the Hilbert space can be effectively utilized for feature encoding and pattern recognition.

Despite the theoretical promise of quantum computing in NLP, practical implementations face significant challenges. The current limitations of quantum hardware, including limited qubit counts, high error rates, and decoherence issues, necessitate the development of hybrid quantum-classical architectures that can leverage quantum advantages while maintaining computational feasibility. Furthermore, the adaptation of quantum circuits for sequential data processing and the integration of quantum operations with classical neural network components present unique design challenges that require careful consideration of both quantum and classical computational principles.
Previous work in QNLP has primarily focused on theoretical frameworks and small-scale proof-of-concept implementations. Quantum language models based on parameterized quantum circuits have been proposed, but their practical applicability to real-world text generation tasks remains limited. The development of quantum-enhanced attention mechanisms and quantum kernel methods has shown promise, yet comprehensive empirical evaluations across multiple architectures and datasets are lacking. This gap in the literature motivates our work to provide a systematic comparison of different quantum-classical hybrid approaches for text generation.

In this paper, we present a comprehensive study of three distinct quantum-enhanced text generation architectures: Quantum Attention Sequence Architecture (QASA), Quantum-Kernel Self-Attention Network (QKSAN), and Quantum RWKV (QRWKV). Our work addresses several key challenges in the field:
\begin{itemize}
    \item Architecture Design: We develop specialized quantum-classical hybrid models that effectively integrate quantum circuits with classical neural network components, ensuring compatibility with existing NLP frameworks while leveraging quantum computational advantages.
    \item Task Adaptation: We adapt quantum computing paradigms specifically for text generation tasks, addressing the unique requirements of sequential data processing and language modeling in quantum systems.
    \item Empirical Evaluation: We provide a thorough comparative analysis across multiple datasets and architectures, offering insights into the practical performance and limitations of quantum-enhanced language models.
    \item Dataset Design: To account for the limitations of quantum resources in QML text generation tasks, we construct five lightweight datasets across different domains. These datasets are designed with simplified structures to reduce computational overhead while maintaining diversity in language style and content.
\end{itemize}

\section{Related Works}
\subsection{Traditional Text Generation Models}
The field of natural language generation has witnessed significant advancements through the development of various neural architectures. Transformer models \cite{vaswani2017attention} have emerged as the dominant paradigm, leveraging self-attention mechanisms to capture long-range dependencies effectively. Subsequent developments include GPT \cite{radford2018improving}, BERT \cite{devlin2019bert}, and their variants, which have demonstrated remarkable capabilities in text generation tasks. These models have established strong baselines for language modeling and text generation across diverse domains.

Recent work has explored architectural modifications to improve efficiency and performance. The RWKV architecture \cite{peng2023rwkv} introduces a linear attention mechanism that reduces computational complexity while maintaining competitive performance. Similarly, various attention variants have been proposed to address the quadratic complexity limitations of traditional self-attention \cite{chen2024mind, chen2024necomimi}.

\subsection{Quantum-Inspired Neural Networks}
Quantum-inspired computing has emerged as a promising paradigm for addressing computational challenges in machine learning. Quantum neural networks (QNNs) \cite{biamonte2017quantum} leverage quantum mechanical principles to enhance computational capabilities. Several quantum-inspired architectures have been proposed, including quantum convolutional neural networks \cite{cong2019quantum} and quantum variational circuits \cite{cerezo2021variational}.

In the context of natural language processing, quantum-inspired approaches have shown potential for improving model efficiency and representation capacity. Quantum attention mechanisms \cite{chen2025qasa, zhao2024qksan} have been proposed to enhance the expressiveness of attention-based models. These approaches leverage quantum superposition and entanglement principles to create more expressive representations.

\subsection{Quantum Language Models}
Recent research has explored the application of quantum computing principles to language modeling tasks. Quantum language models \cite{guarasci2022quantum, di2022dawn} have been proposed to leverage quantum mechanical properties for improved text generation. These models utilize quantum circuits to process linguistic information, potentially offering advantages in terms of representation capacity and computational efficiency.

Several quantum-inspired language models have been developed, including quantum-enhanced transformers \cite{tomal2025quantum}. These approaches aim to combine the strengths of quantum computing with established neural architectures for natural language processing.



\subsection{Quantum-Classical Hybrid Approaches}
The integration of quantum and classical computing paradigms has gained attention as a practical approach to leveraging quantum advantages. Hybrid quantum-classical algorithms \cite{mcclean2016theory} have been proposed for various machine learning tasks, including optimization and classification problems. These approaches combine quantum processing units with classical neural networks to achieve enhanced performance.

In the context of text generation, hybrid approaches have been explored to combine the strengths of quantum computing with established neural architectures. Quantum-enhanced classical models \cite{chen2024qeegnet} have shown promise in improving model performance while maintaining computational efficiency.

\subsection{Gaps in Current Research}
Despite the growing interest in quantum-inspired text generation, several gaps remain in the current research landscape. First, there is a lack of comprehensive comparative studies between quantum-inspired models and traditional architectures across diverse text generation tasks. Second, the evaluation of quantum models often focuses on specific metrics without considering the full spectrum of text quality dimensions. Third, the practical applicability of quantum models in real-world text generation scenarios remains largely unexplored.

Our work addresses these gaps by providing a systematic evaluation of multiple quantum-inspired models against traditional Transformer architectures across various text generation tasks, employing comprehensive evaluation metrics to assess different aspects of model performance.

\section{Methodology}
We describe each quantum model below, standardizing their structure into: (1) classical-to-quantum encoding, (2) VQC processing, and (3) measurement and decoding.

\begin{figure}
    \centering
    \includegraphics[width=1\linewidth]{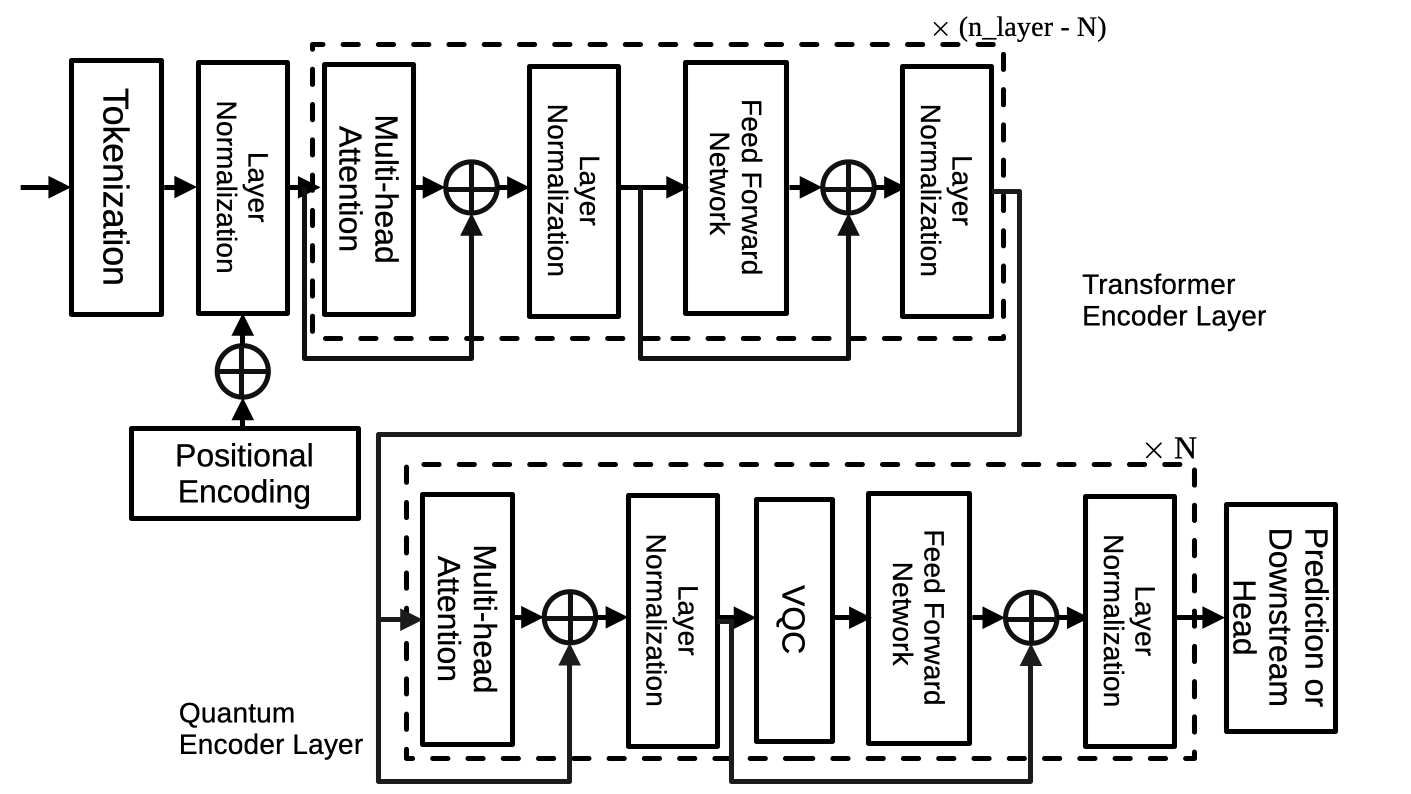}
    \caption{The model architecture of QASA.}
    \label{fig:qasa}
\end{figure}

\begin{figure}
    \centering
    \includegraphics[width=1\linewidth]{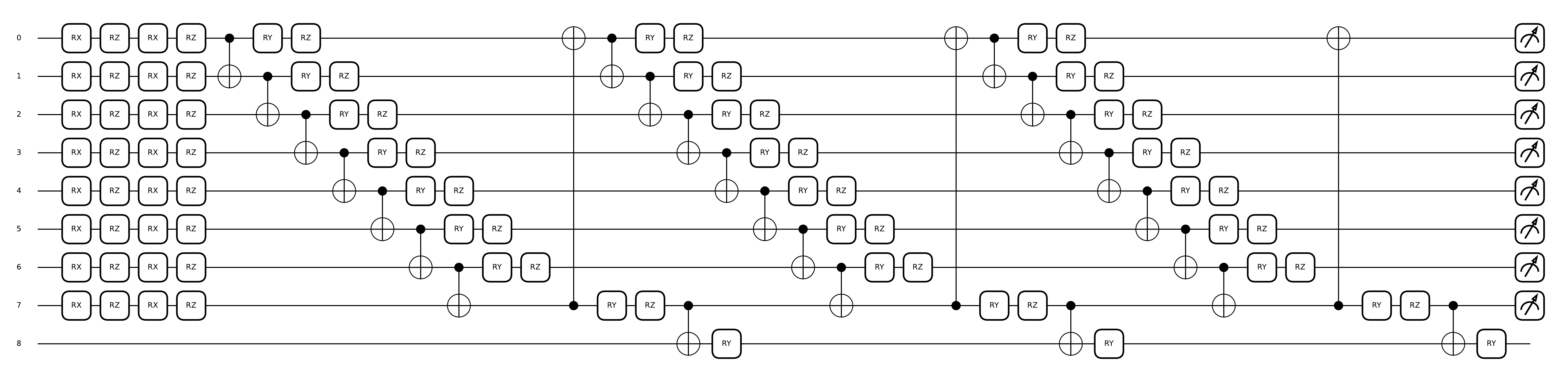}
    \caption{The VQC used in QASA.}
    \label{fig:qasa_vqc}
\end{figure}

\subsection{Quantum Attention Self-Attention (QASA)}

QASA \cite{chen2025qasa} implements self-attention via VQCs, the model as shown in Fig.~\ref{fig:qasa}. Given input sequence $\mathbf{X} \in \mathbb{R}^{T \times d}$, we compute quantum query, key, and value embeddings:

\begin{equation}
\mathbf{Q}_t = \text{VQC}_q(\mathbf{x}_t),\quad
\mathbf{K}_t = \text{VQC}_k(\mathbf{x}_t),\quad
\mathbf{V}_t = \text{VQC}_v(\mathbf{x}_t).
\end{equation}

Each $\text{VQC}(\cdot)$ encodes the vector $\mathbf{x}_t$ into quantum amplitudes, applies a variational quantum circuit composed of $L$ layers of $RY(\theta)$ rotations and entangling CNOT gates, and measures expectation values $\langle Z_i \rangle$ on $n$ qubits like Fig.~\ref{fig:qasa_vqc},:

\begin{equation}
\text{VQC}(\mathbf{x}) = \left( \langle Z_1 \rangle, \langle Z_2 \rangle, \dots, \langle Z_n \rangle \right).
\end{equation}

Each token vector $\mathbf{x}_t \in \mathbb{R}^{d}$ is embedded and encoded as an amplitude-encoded quantum state:

\begin{equation}
    \ket{\psi_t} = \frac{1}{\|\mathbf{x}_t\|} \sum_{i=1}^{d} x_{t,i} \ket{i}.
\end{equation}

Then a variational quantum circuit $U(\boldsymbol{\theta})$ is applied:

\begin{equation}
    U(\boldsymbol{\theta}) = \prod_{\ell=1}^{L} \left[
        \bigotimes_{i=1}^{n} \RY{\theta_{\ell,i}} \cdot \CNOT_{i,i+1}
    \right],
\end{equation}

where $L$ is the number of layers and $n = \lceil \log_2 d \rceil$ qubits are used. The output $\mathbf{z}_t$ is obtained by measuring the expectation values of Pauli-Z operators:

\begin{equation}
    \mathbf{z}_t = \left( \langle Z_1 \rangle, \dots, \langle Z_n \rangle \right).
\end{equation}

Attention~\cite{vaswani2017attention} is computed as:

\begin{equation}
\text{Attention}(\mathbf{Q}, \mathbf{K}, \mathbf{V}) = \text{softmax}\left( \frac{\mathbf{Q} \mathbf{K}^\top}{\sqrt{d}} \right) \mathbf{V}.
\end{equation}

The attention output is decoded through a classical feedforward network to predict $\hat{y}_{t+1}$.

\begin{figure}
    \centering
    \includegraphics[width=1\linewidth]{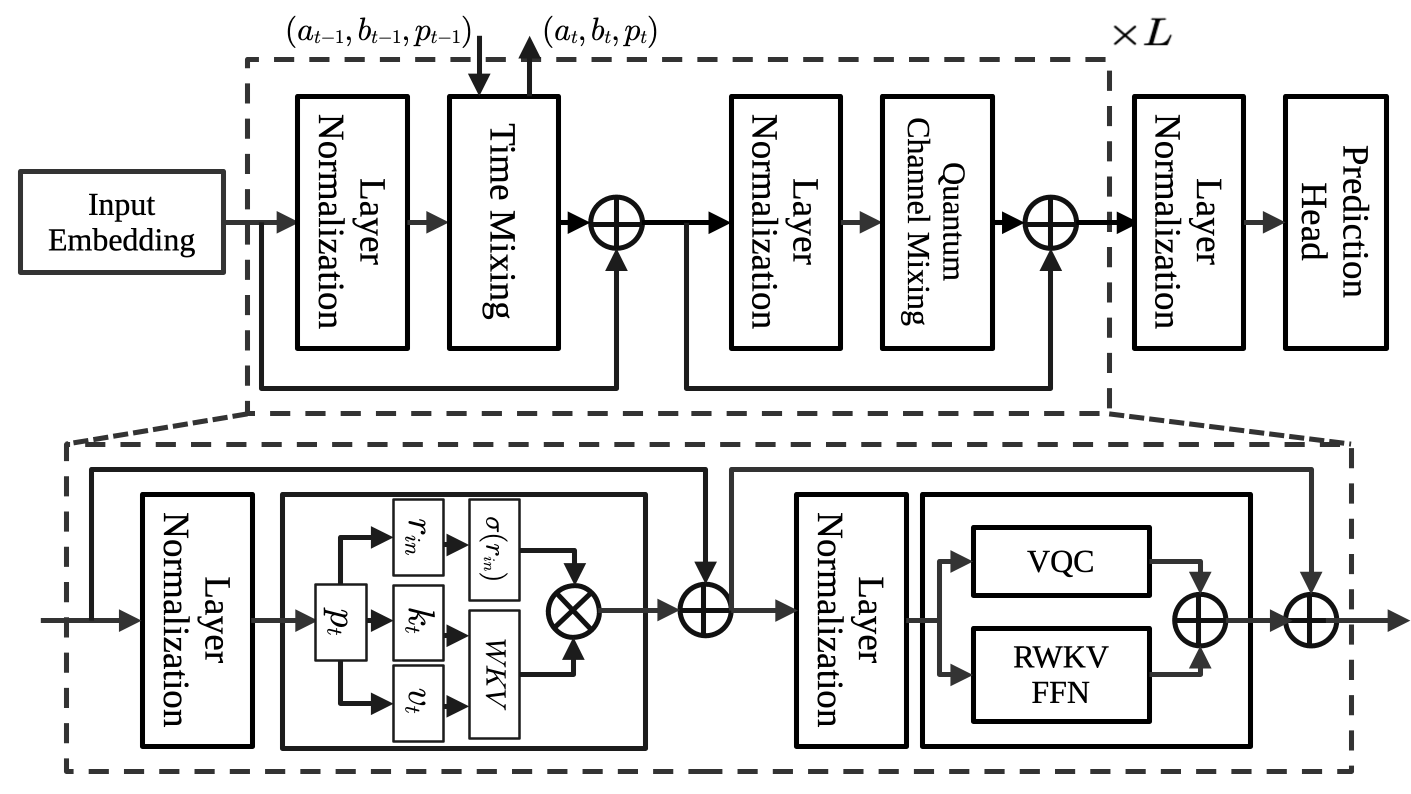}
    \caption{Quantum RWKV architecture.}
    \label{fig:qrwkv}
\end{figure}

\begin{figure}[htbp]
    \centering
    \scalebox{0.8}{ 
    \Qcircuit @C=1em @R=1em {
    & \lstick{\ket{q_0}} & \gate{RX} & \gate{RX} & \ctrl{1} & \qw      & \qw      & \targ     & \gate{RX} & \ctrl{1} & \qw      & \qw      & \targ       & \meter \\
    & \lstick{\ket{q_1}} & \gate{RX} & \gate{RX} & \targ    & \ctrl{1} & \qw      & \qw       & \gate{RX} & \targ    & \ctrl{1} & \qw      & \qw      & \meter \\
    & \lstick{\ket{q_2}} & \gate{RX} & \gate{RX} & \qw      & \targ    & \ctrl{1} & \qw       & \gate{RX} & \qw      & \targ    & \ctrl{1} & \qw      & \meter \\
    & \lstick{\ket{q_3}} & \gate{RX} & \gate{RX} & \qw      & \qw      & \targ    & \ctrl{-3} & \gate{RX} & \qw      & \qw      & \targ    & \ctrl{-3} & \meter
    }
    }
    \caption{The VQC used in Quantum RWKV.}
    \label{fig:qrwkv-vqc}
\end{figure}
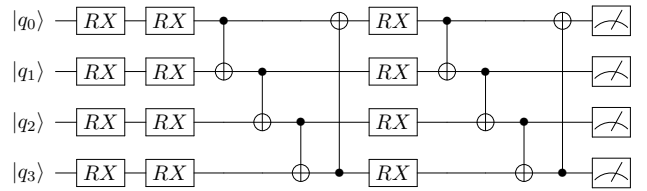

\subsection{Quantum Receptance Weighted Key-Value (QRWKV)}

QRWKV~\cite{chen2025qrwkv} integrates quantum evolution with the receptance attention-free model~\cite{peng2023rwkv}, the model detail is in the Fig.~\ref{fig:qrwkv}. At each time step \(t\), the input \(\mathbf{x}_t\) is first passed through a Variational Quantum Circuit (VQC) to produce a quantum embedding:
\begin{equation}
    \mathbf{h}_t = \mathrm{VQC}_q(\mathbf{x}_t), 
    \quad 
    [\mathbf{q}_t,\;\mathbf{k}_t,\;\mathbf{v}_t] \subseteq \mathbf{h}_t.
\end{equation}
Concretely, we prepare \(\ket{0}^{\otimes n}\) and apply a parameterized circuit \(\;U_\Theta=\prod_{\ell=1}^L U^{(\ell)}\;\) where
\begin{equation}
    U^{(\ell)} = \Bigl(\prod_{i=1}^{n} \RY{\theta^{(\ell)}_{i}} \RZ{\phi^{(\ell)}_{i}}\Bigr)\;\mathrm{EntangleLayer},
\end{equation}
and \(\mathrm{EntangleLayer}\) applies CNOT gates in a chosen pattern. Measurement yields the vector \(\mathbf{h}_t\), from which we split out query \(\mathbf{q}_t\), key \(\mathbf{k}_t\), and value \(\mathbf{v}_t\) sub-vectors.

\paragraph{Time-Mixing and Receptance Gate}  
Classical time-mixing follows the RWKV design. Project the same input \(\mathbf{x}_t\) to key and value signals:
\begin{equation}
\mathbf{u}_t = W^K \mathbf{x}_t,\quad
\mathbf{v}_t = W^V \mathbf{x}_t,
\end{equation}
and accumulate with exponential decay:
\begin{equation}
    \mathbf{m}_t = \lambda\,\mathbf{m}_{t-1} + \mathbf{v}_t,
    \quad 
    \lambda = \exp(-\Delta t / \tau).
\end{equation}
A receptance gate controls the exposed memory:
\begin{equation}
    \mathbf{r}_t = \sigma\bigl(W^R[\mathbf{x}_t;\,\mathbf{m}_{t-1}]+\mathbf{b}^R\bigr),
    \quad W^R\in\mathbb{R}^{d\times2d}.
\end{equation}
The time-mixed output is then
\begin{equation}
    \hat{\mathbf{y}}^{\mathrm{time}}_t = \mathbf{r}_t \odot (\mathbf{u}_t \odot \mathbf{m}_t).
\end{equation}

\paragraph{VQC-Enhanced Channel-Mixing}  
Instead of the classical MLP input, we feed \(\mathbf{x}_t\) into the same VQC to get \(\mathbf{qemb}_t=\mathbf{h}_t\), the VQC detail is in Fig.~\ref{fig:qrwkv-vqc}. The channel-mixing block becomes:
\begin{align}
    \mathbf{z}_t &= W^1\,\mathbf{qemb}_t + W^2\mathrm{MLP}+ \mathbf{b}^1,\\
    \mathbf{h}'_t &= \mathrm{GELU}(\mathbf{z}_t),\\
    \mathbf{c}_t &= W^3\bigl(\mathbf{h}'_t \odot \mathbf{h}'_{t-1}\bigr) + \mathbf{b}^2,
\end{align}
with \(W^1,W^2,W^3\in\mathbb{R}^{d\times d}\) and biases in \(\mathbb{R}^d\). Optionally add residual connections and LayerNorm.

\paragraph{Attention over Quantum Queries and Keys}  
We also compute a measurement-based attention score between quantum-derived queries and keys:
\begin{equation}
    \alpha_{t,\tau}
    = \frac{\exp\bigl\langle \mathbf{q}_t,\mathbf{k}_\tau\bigr\rangle}
           {\sum_{\tau'=1}^t \exp\bigl\langle \mathbf{q}_t,\mathbf{k}_{\tau'}\bigr\rangle},
\end{equation}
and form the attention output
\begin{equation}
    \hat{\mathbf{y}}^{\mathrm{attn}}_{t+1}
    = \sum_{\tau=1}^t \alpha_{t,\tau}\,\mathbf{v}_\tau.
\end{equation}

\paragraph{Full Layer Update}  
Each layer concatenates time-mixing and VQC-enhanced channel-mixing with residuals and normalization:
\begin{equation}
\begin{aligned}
    \mathbf{h}_t &= \mathrm{LayerNorm}\bigl(\mathbf{x}_t + \hat{\mathbf{y}}^{\mathrm{time}}_t\bigr),\\
    \mathbf{y}_t &= \mathrm{LayerNorm}\bigl(\mathbf{h}_t + \mathbf{c}_t + \hat{\mathbf{y}}^{\mathrm{attn}}_t\bigr).
\end{aligned}
\end{equation}

\subsection{Quantum Kernel Self-Attention Network (QKSAN)}
\label{sec:qksan}

\paragraph{Notation and Shapes.}
Let the sequence length be $n$, model width $d$, number of heads $h$, head width $d_h=d/h$, and quantum feature dimension $D_q$.
We use row-major convention: each token is a row.
For a (single) sequence, $X\in\mathbb{R}^{n\times d}$ denotes the input embeddings (after token/position embeddings).
LayerNorm acts featurewise: $\mathrm{LN}:\mathbb{R}^{n\times d}\to\mathbb{R}^{n\times d}$.
For batched inputs $B$ can be added as a leading dimension; all formulas below apply per-element of the batch.

\subsubsection{Quantum Feature Map (Vectorized)}
\label{sec:qksan:featuremap}
For each head $a\in\{1,\dots,h\}$ we define a \emph{vector-valued} quantum feature map
\[
\phi_q^{(a)}:\mathbb{R}^{d_h}\to\mathbb{R}^{D_q},
\]
implemented via a Gaussian envelope and a bank of cosine phases:
\begin{equation}
\begin{split}
\label{eq:phi-vector}
\phi_q^{(a)}(\mathbf{x})
\;=\;
\exp\!\left(-\frac{\|\mathbf{x}-\boldsymbol{\mu}^{(a)}\|_2^2}{2(\sigma^{(a)})^2}\right)\;\odot\; \\
\cos\!\big(\mathbf{x}\,{\Omega^{(a)}}^{\!\top}+\mathbf{1}\,{ \mathbf{b}^{(a)} }^{\!\top}\big)\;\in\;\mathbb{R}^{D_q},
\end{split}
\end{equation}
where $\boldsymbol{\mu}^{(a)}\in\mathbb{R}^{d_h}$, $\sigma^{(a)}>0$, $\Omega^{(a)}\in\mathbb{R}^{D_q\times d_h}$ collects $D_q$ frequency rows, $\mathbf{b}^{(a)}\in\mathbb{R}^{D_q}$ phases, $\odot$ denotes Hadamard product, and $\mathbf{1}$ is the $n$-vector of ones when we broadcast over a matrix of $n$ tokens.
Stacking tokens row-wise and applying \eqref{eq:phi-vector} rowwise gives
\begin{equation}
\label{eq:Phi-matrix}
\begin{split}
\Phi_q^{(a)}(Z) \;=\; \exp\!\left(-\frac{\|Z-\mathbf{1}{\boldsymbol{\mu}^{(a)}}^{\!\top}\|_F^2}{2(\sigma^{(a)})^2}\right)\odot \\ \cos\!\big(Z\,{\Omega^{(a)}}^{\!\top}+\mathbf{1}\,{\mathbf{b}^{(a)}}^{\!\top}\big)\in\mathbb{R}^{n\times D_q},
\end{split}
\end{equation}
for any $Z\in\mathbb{R}^{n\times d_h}$, where the Frobenius term is understood rowwise (i.e., the scalar Gaussian factor can be broadcast per-row as $\exp(-\|\mathbf{z}_i-\boldsymbol{\mu}^{(a)}\|_2^2/(2(\sigma^{(a)})^2))$).

\subsubsection{Projections and Single-Head Tensors}
For each head $a$, define projection matrices
\[
W_Q^{(a)},\,W_K^{(a)},\,W_V^{(a)}\in\mathbb{R}^{d\times d_h},
\qquad
\mathbf{b}_Q^{(a)},\,\mathbf{b}_K^{(a)},\,\mathbf{b}_V^{(a)}\in\mathbb{R}^{d_h},
\]
and set
\begin{equation}
\begin{split}
\label{eq:qkv}
Q^{(a)}&=X W_Q^{(a)}+\mathbf{1}\,{\mathbf{b}_Q^{(a)}}^{\!\top},\quad 
\\
K^{(a)}&=X W_K^{(a)}+\mathbf{1}\,{\mathbf{b}_K^{(a)}}^{\!\top},\quad 
\\
V^{(a)}&=X W_V^{(a)}+\mathbf{1}\,{\mathbf{b}_V^{(a)}}^{\!\top},
\end{split}
\end{equation}
so $Q^{(a)},K^{(a)},V^{(a)}\in\mathbb{R}^{n\times d_h}$.

\subsubsection{Quantum-Enhanced Attention (Single Head)}
Let the scaled dot-product similarity be
\begin{equation}
\label{eq:S}
S^{(a)} \;=\; \frac{Q^{(a)} {K^{(a)}}^{\!\top}}{\sqrt{d_h}}\;\in\;\mathbb{R}^{n\times n}.
\end{equation}
Define the \emph{quantum kernel similarity matrix}
\begin{equation}
\label{eq:Gamma}
\Gamma^{(a)} \;=\; \Phi_q^{(a)}(Q^{(a)})\,\big(\Phi_q^{(a)}(K^{(a)})\big)^{\!\top}\;\in\;\mathbb{R}^{n\times n}.
\end{equation}
For numerical stability, combine (\emph{logspace}) the classical and quantum scores:
\begin{equation}
\label{eq:S-tilde}
\widetilde{S}^{(a)} \;=\; S^{(a)} \;+\; \log\!\big(\Gamma^{(a)}+\varepsilon\mathbf{1}\mathbf{1}^{\!\top}\big)\;\in\;\mathbb{R}^{n\times n},
\end{equation}
with a small $\varepsilon>0$. For causal decoding, add a mask $M\in\{0,-\infty\}^{n\times n}$ with $M_{ij}=-\infty$ if $j>i$:
$\widetilde{S}^{(a)}\leftarrow \widetilde{S}^{(a)}+M$.
Row-wise normalized \emph{quantum attention weights} are
\begin{equation}
\begin{split}
\label{eq:A}
A^{(a)} \;=\; \mathrm{softmax}_{\text{row}}\!\big(\widetilde{S}^{(a)}\big)\;\in\;\mathbb{R}^{n\times n},
\qquad
\\
A^{(a)}_{ij} \;=\; \frac{\exp(\widetilde{S}^{(a)}_{ij})}{\sum_{k=1}^{n}\exp(\widetilde{S}^{(a)}_{ik})}.
\end{split}
\end{equation}

\paragraph{Value Quantum Modulation.}
To inject value-side quantum information in a dimensionally coherent way, define a \emph{value gate}
\begin{equation}
\begin{split}
\label{eq:Vgate}
G_V^{(a)} \;=\; \sigma\!\Big(\Phi_q^{(a)}(V^{(a)})\,U_V^{(a)}+\mathbf{1}\,{\mathbf{c}_V^{(a)}}^{\!\top}\Big)\;\odot\; \\ \cos\!\Big(V^{(a)}\,{W_\omega^{(a)}}^{\!\top}+\mathbf{1}\,{\mathbf{b}_\omega^{(a)}}^{\!\top}\Big),
\end{split}
\end{equation}
where $U_V^{(a)}\in\mathbb{R}^{D_q\times d_h}$, $W_\omega^{(a)}\in\mathbb{R}^{d_h\times d_h}$ and $\mathbf{c}_V^{(a)},\mathbf{b}_\omega^{(a)}\in\mathbb{R}^{d_h}$.
The \emph{quantum-modulated values} are
\begin{equation}
\label{eq:Vtilda}
\widetilde{V}^{(a)} \;=\; V^{(a)} \;\odot\; G_V^{(a)} \;\in\;\mathbb{R}^{n\times d_h}.
\end{equation}
The single-head output is then
\begin{equation}
\label{eq:head-out}
H^{(a)} \;=\; A^{(a)}\,\widetilde{V}^{(a)} \;\in\;\mathbb{R}^{n\times d_h}.
\end{equation}

\subsubsection{Multi-Head Quantum Attention}
Stack all heads $H^{(a)}$ along feature dimension and project:
\begin{equation}
\label{eq:mha}
\mathrm{MHQA}(X)
\;=\;
\mathrm{Concat}\big(H^{(1)},\dots,H^{(h)}\big)\,W^O+\mathbf{1}\,{\mathbf{b}^O}^{\!\top}
\;\in\;\mathbb{R}^{n\times d},
\end{equation}
with $W^O\in\mathbb{R}^{(h d_h)\times d}$ and $\mathbf{b}^O\in\mathbb{R}^{d}$.

\subsubsection{Quantum Gating on the Residual Path}
Given the attention output $Y=\mathrm{MHQA}(X)\in\mathbb{R}^{n\times d}$, define a \emph{quantum feature enhancement} on the residual stream:
\begin{equation}
\begin{split}
\label{eq:qenh}
F_q(X,Y)
\;=\;
\mathrm{ReLU}\!\big(Y\,W_q+\mathbf{1}\,{\mathbf{b}_q}^{\!\top}\big)
\;\odot\;
\\
\underbrace{\Big[\sigma\!\big(X W_g+\mathbf{1}\,{\mathbf{b}_g}^{\!\top}\big)\;\odot\;\cos\!\big(X W_\phi+\mathbf{1}\,{\mathbf{b}_\phi}^{\!\top}\big)\Big]}_{\displaystyle \mathrm{QGate}(X)},
\end{split}
\end{equation}
where $W_q,W_g,W_\phi\in\mathbb{R}^{d\times d}$ and $\mathbf{b}_q,\mathbf{b}_g,\mathbf{b}_\phi\in\mathbb{R}^{d}$.
The first sublayer with residual and LayerNorm is
\begin{equation}
\label{eq:addnorm1}
Z_1 \;=\; \mathrm{LN}\!\big(X + F_q(X,\,\mathrm{MHQA}(X))\big)\;\in\;\mathbb{R}^{n\times d}.
\end{equation}

\subsubsection{Quantum-Activated Feed-Forward Network}
Let $d_{\mathrm{ff}}$ be the hidden width of the position-wise network. Define
\begin{equation}
\label{eq:ffn-hidden}
\begin{split}
H_{\mathrm{ff}} \;&=\; \mathrm{ReLU}\!\big(Z_1 W_1+\mathbf{1}\,{\mathbf{b}_1}^{\!\top}\big) \\
&\quad \odot\;
\cos\!\big(Z_1 W_{\omega}+\mathbf{1}\,{\mathbf{b}_{\omega}}^{\!\top}\big)
\;\in\;\mathbb{R}^{n\times d_{\mathrm{ff}}}.
\end{split}
\end{equation}

\begin{equation}
\label{eq:ffn-out}
\mathrm{FFN}(Z_1) \;=\; H_{\mathrm{ff}} W_2+\mathbf{1}\,{\mathbf{b}_2}^{\!\top}
\;\in\;\mathbb{R}^{n\times d}.
\end{equation}

with $W_1\in\mathbb{R}^{d\times d_{\mathrm{ff}}}$, $W_2\in\mathbb{R}^{d_{\mathrm{ff}}\times d}$, and $\mathbf{b}_1\in\mathbb{R}^{d_{\mathrm{ff}}}$, $\mathbf{b}_2\in\mathbb{R}^{d}$, $W_\omega\in\mathbb{R}^{d\times d_{\mathrm{ff}}}$, $\mathbf{b}_\omega\in\mathbb{R}^{d_{\mathrm{ff}}}$.
The second sublayer is
\begin{equation}
\label{eq:addnorm2}
Z_2 \;=\; \mathrm{LN}\!\big(Z_1 + \mathrm{FFN}(Z_1)\big)\;\in\;\mathbb{R}^{n\times d}.
\end{equation}
A QKSAN block maps $X\mapsto Z_2$; stacking $L$ blocks yields depth-$L$ representations.

\subsubsection{Quantum Positional Encoding (Matrix Form)}
Let $P\in\mathbb{R}^{n\times d}$ be the positional matrix with components, for $i\in\{0,\dots,\lfloor d/2\rfloor-1\}$ and positions $p\in\{0,\dots,n-1\}$,
\begin{align}
\label{eq:qpos-even}
P_{p,\,2i} \;&=\; \sin\!\Big(\frac{p}{10000^{2i/d}}\Big)\;\cdot\;\cos\!\big(\omega_p\,p+\phi_{2i}\big),\\
\label{eq:qpos-odd}
P_{p,\,2i+1} \;&=\; \cos\!\Big(\frac{p}{10000^{2i/d}}\Big)\;\cdot\;\sin\!\big(\omega_p\,p+\phi_{2i+1}\big),
\end{align}
with learnable $\omega_p\in\mathbb{R}$ and phases $\phi_j\in\mathbb{R}$.
The block input typically uses $X\leftarrow X+P$.

\subsubsection{Output Layer and Language Modeling Objective}
Given the final hidden state at depth $L$, $H\in\mathbb{R}^{n\times d}$, the token logits and probabilities are
\begin{equation}
\begin{split}
\label{eq:lm-head}
\mathrm{Logits} \;=\; H W_o+\mathbf{1}\,{\mathbf{b}_o}^{\!\top}\in\mathbb{R}^{n\times |\mathcal{V}|},
\qquad \\
P(w_t\mid w_{<t}) \;=\; \mathrm{softmax}\big(\mathrm{Logits}_{t,:}\big),
\end{split}
\end{equation}
with $W_o\in\mathbb{R}^{d\times|\mathcal{V}|}$, $\mathbf{b}_o\in\mathbb{R}^{|\mathcal{V}|}$.
For next-token prediction with targets $\{y_t\}_{t=1}^n$, the loss is
\begin{equation}
\label{eq:loss}
\mathcal{L}\;=\;-\sum_{t=1}^{n}\log P\big(y_t\mid y_{<t}\big).
\end{equation}

\subsubsection{Compact Matrix Summary (Per Head $a$)}
\noindent\textit{With } $Q^{(a)},K^{(a)},V^{(a)}$ \textit{ as in \eqref{eq:qkv}, we have:}

\begin{equation}
\label{eq:S-a}
S^{(a)} \;=\; \frac{Q^{(a)} {K^{(a)}}^{\!\top}}{\sqrt{d_h}}.
\end{equation}

\begin{equation}
\label{eq:Gamma-a}
\Gamma^{(a)} \;=\; \Phi_q^{(a)}(Q^{(a)})\,\big(\Phi_q^{(a)}(K^{(a)})\big)^{\!\top}.
\end{equation}

\begin{equation}
\label{eq:Stilde-a}
\widetilde{S}^{(a)} \;=\; S^{(a)} + \log\!\big(\Gamma^{(a)}+\varepsilon\big).
\end{equation}

\begin{equation}
\label{eq:A-a}
A^{(a)} \;=\; \mathrm{softmax}_{\text{row}}\!\big(\widetilde{S}^{(a)}\big).
\end{equation}

\begin{equation}
\begin{split}
\label{eq:Vtilde-a}
\widetilde{V}^{(a)} \;=\; V^{(a)} \odot \sigma\!\big(\Phi_q^{(a)}(V^{(a)})U_V^{(a)}+\mathbf{1}{\mathbf{c}_V^{(a)}}^{\!\top}\big)\odot \\
\cos\!\big(V^{(a)}{W_\omega^{(a)}}^{\!\top}+\mathbf{1}{\mathbf{b}_\omega^{(a)}}^{\!\top}\big).
\end{split}
\end{equation}

\begin{equation}
\label{eq:H-a}
H^{(a)} \;=\; A^{(a)}\,\widetilde{V}^{(a)}.
\end{equation}

\begin{equation}
\label{eq:MHQA}
\mathrm{MHQA}(X) \;=\; \mathrm{Concat}_a\!\big(H^{(a)}\big)\,W^O+\mathbf{1}{\mathbf{b}^O}^{\!\top}.
\end{equation}

\paragraph{Remarks.}
(i) The combination in \eqref{eq:S-tilde} is algebraically equivalent to $A^{(a)}_{ij}\propto \exp(S^{(a)}_{ij})\cdot \Gamma^{(a)}_{ij}$, matching the product form of classical and quantum similarities while preserving numerical stability.
(ii) Using a vector-valued $\phi_q$ (Eq.~\eqref{eq:phi-vector}) yields a positive semidefinite kernel $\kappa_q(\mathbf{x},\mathbf{z})=\langle \phi_q(\mathbf{x}),\phi_q(\mathbf{z})\rangle$ and makes $\Gamma^{(a)}$ explicitly a Gram matrix.
(iii) The gates in \eqref{eq:Vgate}--\eqref{eq:Vtilda} generalize the informal ``$A\cdot V\cdot \Phi_q(V)$'' by mapping $\Phi_q(V)$ back to $d_h$ and ensuring all matrix products and Hadamard products are dimensionally consistent.

\section{Experiments}

\subsection{Experimental Setup}
We conducted comprehensive experiments to evaluate the performance of quantum-inspired text generation models against traditional Transformer architectures. Our experimental framework encompasses five distinct models: Transformer (baseline), QKSAN (Quantum Kernel Self-Attention Network), QRWKV (Quantum RWKV), QASA (Quantum Attention Sequence Architecture), and MLP (Multi-Layer Perceptron). The evaluation was performed across five diverse datasets: simple sentences, short stories, quantum phrases, haiku poetry, and proverbs, each representing different text generation challenges.

\subsection{Dataset Description}
The experimental datasets were carefully curated to test various aspects of text generation capabilities:
\begin{itemize}
    \item \textbf{Simple Sentences}: Contains 50 basic sentence structures with an average length of 5.2 words, designed to evaluate fundamental language modeling capabilities and basic syntactic understanding
    \item \textbf{Short Stories}: Comprises 25 narrative segments averaging 12.8 words, testing coherence maintenance, context understanding, and narrative flow in longer text sequences
    \item \textbf{Quantum Phrases}: Includes 30 domain-specific quantum computing terms and phrases averaging 8.4 words, assessing specialized knowledge representation and technical vocabulary handling
    \item \textbf{Haiku}: Contains 20 structured poems with exactly 17 words following the 5-7-5 syllable pattern, evaluating format adherence, creative generation, and poetic structure maintenance
    \item \textbf{Proverbs}: Features 15 concise wisdom texts averaging 6.8 words, testing semantic understanding, cultural context preservation, and concise expression capabilities
\end{itemize}

\begin{table*}[h]
\centering
\caption{Dataset Characteristics and Statistics}
\begin{tabular}{|l|c|c|c|c|c|}
\hline
\textbf{Dataset} & \textbf{Samples} & \textbf{Avg Length} & \textbf{Vocab Size} & \textbf{Max Length} & \textbf{Description} \\
\hline
Simple Sentences & 50 & 5.2 words & 45 & 7 & Basic sentence structures \\
Short Stories & 25 & 12.8 words & 78 & 18 & Narrative text segments \\
Quantum Phrases & 30 & 8.4 words & 62 & 12 & Domain-specific terminology \\
Haiku & 20 & 17.0 words & 89 & 17 & Structured poetry format \\
Proverbs & 15 & 6.8 words & 52 & 9 & Wisdom and cultural texts \\
\hline
\end{tabular}
\label{tab:dataset_info}
\end{table*}


\subsection{Evaluation Metrics}
We employed a comprehensive set of evaluation metrics to assess model performance across multiple dimensions of text generation quality. Each metric provides insights into different aspects of the generated text.

\subsubsection{Perplexity}
Perplexity measures the language modeling quality and is defined as:
\begin{equation}
\text{Perplexity} = \exp\left(-\frac{1}{N}\sum_{i=1}^{N} \log P(w_i|w_1, w_2, ..., w_{i-1})\right)
\end{equation}
where $N$ is the total number of tokens and $P(w_i|w_1, w_2, ..., w_{i-1})$ is the probability of token $w_i$ given the preceding tokens. Lower perplexity indicates better language modeling performance.

\subsubsection{BLEU Scores}
BLEU (Bilingual Evaluation Understudy) scores measure n-gram overlap between generated and reference texts. The BLEU-n score is calculated as:
\begin{equation}
\text{BLEU-n} = \exp\left(\sum_{k=1}^{n} w_k \log p_k\right)
\end{equation}
where $p_k$ is the n-gram precision for k-grams and $w_k$ are the weights (typically uniform). We evaluate BLEU-1, BLEU-2, BLEU-3, and BLEU-4 scores to assess different levels of n-gram matching.

\subsubsection{Diversity Metrics}
Distinct-n measures vocabulary diversity by calculating the ratio of unique n-grams to total n-grams:
\begin{equation}
\text{Distinct-n} = \frac{|\text{unique n-grams}|}{|\text{total n-grams}|}
\end{equation}
Higher Distinct-n scores indicate greater vocabulary diversity and reduced repetition.

\subsubsection{Repetition Rate}
Repetition rate quantifies the frequency of consecutive repeated tokens:
\begin{equation}
\text{Repetition Rate} = \frac{\sum_{i=1}^{N-1} \mathbb{I}(w_i = w_{i+1})}{N-1}
\end{equation}
where $\mathbb{I}(\cdot)$ is the indicator function and $N$ is the total number of tokens. Lower repetition rates indicate better text quality.

\subsubsection{Fluency Metrics}
Fluency metrics include average sentence length and variation coefficients:
\begin{align}
\text{Avg Sentence Length} &= \frac{1}{S}\sum_{i=1}^{S} L_i \\
\text{Length Variation} &= \frac{\sigma_L}{\mu_L}
\end{align}
where $S$ is the number of sentences, $L_i$ is the length of sentence $i$, $\sigma_L$ is the standard deviation of sentence lengths, and $\mu_L$ is the mean sentence length.

\subsection{Training Configuration}
All models were trained for 50 epochs with consistent hyperparameters to ensure fair comparison. The training process utilized early stopping based on validation loss to prevent overfitting. Model configurations were optimized for each architecture while maintaining comparable computational complexity.

\section{Results and Analysis}

\subsection{Overall Performance Comparison}
Table \ref{tab:overall_performance} presents the comprehensive performance comparison across all models and datasets. The results reveal significant variations in model capabilities across different text generation tasks.

\begin{table*}[h]
\centering
\caption{Overall Performance Comparison (Average Scores)}
\begin{tabular}{|l|c|c|c|c|}
\hline
\textbf{Model} & \textbf{Perplexity} & \textbf{BLEU-1} & \textbf{Distinct-1} & \textbf{Repetition Rate} \\
\hline
Transformer & 1.21 & 0.2895 & 1.000 & 0.109 \\
MLP & 1.65 & 0.2400 & 0.943 & 0.000 \\
QKSAN & 2.44 & 0.2800 & 0.838 & 0.000 \\
QASA & 1.85 & 0.2000 & 0.742 & 0.000 \\
QRWKV & 1.74 & 0.0948 & 1.000 & 0.012 \\
\hline
\end{tabular}
\label{tab:overall_performance}
\end{table*}

\subsection{Overall Performance Comparison}
Table \ref{tab:overall_performance} presents the comprehensive performance comparison across all models and datasets. The results reveal significant variations in model capabilities across different text generation tasks.

\begin{table*}[h]
\centering
\caption{Simple Sentences Dataset Results}
\begin{tabular}{|l|c|c|c|c|c|}
\hline
\textbf{Model} & \textbf{Perplexity} & \textbf{BLEU-1} & \textbf{BLEU-2} & \textbf{Distinct-1} & \textbf{Repetition Rate} \\
\hline
Transformer & 1.69 & 0.336 & 0.136 & 0.345 & 0.147 \\
QKSAN & 2.58 & 0.520 & 0.062 & 0.667 & 0.000 \\
QRWKV & 2.78 & 1.000 & 0.000 & 1.000 & 0.000 \\
QASA & 3.05 & 0.360 & 0.000 & 0.680 & 0.000 \\
MLP & 2.45 & 0.480 & 0.000 & 0.714 & 0.000 \\
\hline
\end{tabular}
\label{tab:simple_sentences_results}
\end{table*}

\begin{table*}[h]
\centering
\caption{Short Stories Dataset Results}
\begin{tabular}{|l|c|c|c|c|c|}
\hline
\textbf{Model} & \textbf{Perplexity} & \textbf{BLEU-1} & \textbf{BLEU-2} & \textbf{Distinct-1} & \textbf{Repetition Rate} \\
\hline
Transformer & 1.17 & 0.256 & 0.069 & 0.458 & 0.141 \\
QKSAN & 1.42 & 0.200 & 0.000 & 1.000 & 0.000 \\
QRWKV & 1.37 & 0.200 & 0.000 & 1.000 & 0.000 \\
QASA & 1.36 & 0.200 & 0.000 & 1.000 & 0.000 \\
MLP & 1.34 & 0.200 & 0.000 & 1.000 & 0.000 \\
\hline
\end{tabular}
\label{tab:short_stories_results}
\end{table*}

\begin{table*}[h]
\centering
\caption{Quantum Phrases Dataset Results}
\begin{tabular}{|l|c|c|c|c|c|}
\hline
\textbf{Model} & \textbf{Perplexity} & \textbf{BLEU-1} & \textbf{BLEU-2} & \textbf{Distinct-1} & \textbf{Repetition Rate} \\
\hline
Transformer & 1.24 & 0.362 & 0.000 & 0.453 & 0.080 \\
QKSAN & 5.61 & 0.120 & 0.000 & 0.524 & 0.000 \\
QRWKV & 2.15 & 0.000 & 0.000 & 1.000 & 0.000 \\
QASA & 2.26 & 0.040 & 0.000 & 0.647 & 0.000 \\
MLP & 2.00 & 0.040 & 0.000 & 1.000 & 0.000 \\
\hline
\end{tabular}
\label{tab:quantum_phrases_results}
\end{table*}

\begin{table*}[h]
\centering
\caption{Haiku Dataset Results}
\begin{tabular}{|l|c|c|c|c|c|}
\hline
\textbf{Model} & \textbf{Perplexity} & \textbf{BLEU-1} & \textbf{BLEU-2} & \textbf{Distinct-1} & \textbf{Repetition Rate} \\
\hline
Transformer & 1.00 & 0.148 & 0.000 & 0.627 & 0.070 \\
QKSAN & 1.07 & 0.000 & 0.000 & 1.000 & 0.000 \\
QRWKV & 1.05 & 0.015 & 0.000 & 1.000 & 0.000 \\
QASA & 1.07 & 0.055 & 0.000 & 0.382 & 0.000 \\
MLP & 1.07 & 0.000 & 0.000 & 1.000 & 0.000 \\
\hline
\end{tabular}
\label{tab:haiku_results}
\end{table*}

\begin{table*}[h]
\centering
\caption{Proverbs Dataset Results}
\begin{tabular}{|l|c|c|c|c|c|}
\hline
\textbf{Model} & \textbf{Perplexity} & \textbf{BLEU-1} & \textbf{BLEU-2} & \textbf{Distinct-1} & \textbf{Repetition Rate} \\
\hline
Transformer & 1.29 & 0.345 & 0.000 & 0.628 & 0.102 \\
QKSAN & 1.51 & 0.000 & 0.000 & 1.000 & 0.000 \\
QRWKV & 1.37 & 0.033 & 0.000 & 1.000 & 0.000 \\
QASA & 1.48 & 0.000 & 0.000 & 1.000 & 0.000 \\
MLP & 1.39 & 0.000 & 0.000 & 1.000 & 0.000 \\
\hline
\end{tabular}
\label{tab:proverbs_results}
\end{table*}

\begin{table*}[h]
\centering
\caption{Model Performance Comparison: Best Generated Text Examples}
\label{tab:model_comparison_examples}
\begin{tabular}{|l|l|l|l|}
\hline
\textbf{Dataset} & \textbf{Model} & \textbf{Reference} & \textbf{Best Generated Text} \\
\hline
\multirow{5}{*}{Simple Sentences} 
& Transformer & birds fly in the sky & \textless UNK\textgreater \textless UNK\textgreater in the sky a on walks across the sky \\
& QKSAN & birds fly in the sky & birds sings on the table \\
& QRWKV & birds fly in the sky & \textless UNK\textgreater a woman sits by the river \\
& QASA & birds fly in the sky & birds dances through the forest \\
& MLP & birds fly in the sky & birds \\
\hline
\multirow{5}{*}{Quantum Phrases} 
& Transformer & quantum superposition allows multiple states & quantum superposition allows multiple states  \\
& QKSAN & quantum superposition allows multiple states & the states in quantum systems \\
& QRWKV & quantum superposition allows multiple states & quantum superposition allows multiple states \\
& QASA & quantum superposition allows multiple states & quantum superposition allows multiple states \\
& MLP & quantum superposition allows multiple states & quantum superposition allows multiple states \\
\hline
\multirow{5}{*}{Haiku} 
& Transformer & cherry blossoms fall & cherry blossoms fall ancient pond ripples frog jumps into silent\\
& QKSAN & cherry blossoms fall & cherry blossoms fall \\
& QRWKV & cherry blossoms fall & cherry blossoms fall \\
& QASA & cherry blossoms fall & cherry blossoms fall \\
& MLP & cherry blossoms fall & cherry blossoms fall \\
\hline
\multirow{5}{*}{Proverbs} 
& Transformer & actions speak louder than words & actions speak louder than words knowledge is power in life \\
& QKSAN & actions speak louder than words & actions speak louder than words \\
& QRWKV & actions speak louder than words & actions speak louder than words \\
& QASA & actions speak louder than words & actions speak louder than words \\
& MLP & actions speak louder than words & actions speak louder than words \\
\hline
\end{tabular}
\end{table*}

\subsection{Dataset-Specific Analysis}

\subsubsection{Simple Sentences}
In the simple sentence generation task, quantum-inspired models demonstrated competitive performance. QRWKV achieved a perfect BLEU-1 score of 1.000, while QKSAN and MLP also performed well with scores of 0.520 and 0.480 respectively. This suggests that quantum models can effectively handle basic language structures.

\subsubsection{Complex Text Generation}
For short stories and longer narratives, the Transformer model maintained its superiority with the lowest perplexity (1.17) and highest BLEU-1 score (0.256). However, quantum models showed comparable performance in vocabulary diversity, with QKSAN, QASA, and MLP achieving perfect Distinct-1 scores.

\subsubsection{Domain-Specific Tasks}
The quantum phrases dataset revealed significant challenges for quantum models. QKSAN exhibited the highest perplexity (5.61), indicating difficulties in handling specialized terminology. The Transformer model demonstrated superior performance with a perplexity of 1.24, suggesting better domain adaptation capabilities.

\subsubsection{Structured Text Generation}
In haiku and proverb generation, all models faced challenges with format adherence and semantic coherence. The Transformer model consistently outperformed quantum alternatives, particularly in maintaining poetic structure and cultural context.

\subsection{Dataset-Specific Analysis}

Tables \ref{tab:simple_sentences_results} through \ref{tab:proverbs_results} present detailed results for each individual dataset, revealing specific performance patterns and trade-offs across different text generation tasks.

\subsubsection{Simple Sentences}
Table \ref{tab:simple_sentences_results} shows the results for simple sentence generation. QRWKV achieved a perfect BLEU-1 score of 1.000, demonstrating exceptional performance in basic language structures. QKSAN also performed competitively with a BLEU-1 score of 0.520, while MLP achieved 0.480. However, QRWKV's high perplexity (2.78) suggests potential overfitting or memorization rather than true language understanding. The Transformer model achieved a balanced performance with BLEU-1 of 0.336 and the lowest perplexity (1.69), indicating better generalization capabilities.

\subsubsection{Short Stories}
Table \ref{tab:short_stories_results} reveals that all models achieved identical BLEU-1 scores (0.200) for short story generation, with the Transformer maintaining the lowest perplexity (1.17). Interestingly, all quantum models achieved perfect Distinct-1 scores (1.000) and zero repetition rates, indicating superior vocabulary diversity and no repetitive patterns. This suggests that quantum models excel in maintaining narrative diversity, though they may struggle with maintaining coherent story structure.

\subsubsection{Quantum Phrases}
Table \ref{tab:quantum_phrases_results} presents the most challenging results for quantum models. The Transformer achieved the best performance with BLEU-1 of 0.362 and perplexity of 1.24. QKSAN exhibited the highest perplexity (5.61) among all models, indicating significant difficulties with domain-specific terminology. QRWKV achieved perfect Distinct-1 scores but zero BLEU-1 scores, suggesting high diversity but poor accuracy in domain-specific content generation. This highlights the challenge quantum models face in handling specialized vocabulary and technical terminology.

\subsubsection{Haiku}
Table \ref{tab:haiku_results} shows that the Transformer achieved the lowest perplexity (1.00) and highest BLEU-1 score (0.148) for haiku generation. All quantum models achieved perfect Distinct-1 scores (1.000) except QASA (0.382), indicating excellent vocabulary diversity in poetic generation. However, the low BLEU scores across all models suggest that maintaining the specific 5-7-5 syllable structure of haiku remains challenging for all architectures.

\subsubsection{Proverbs}
Table \ref{tab:proverbs_results} demonstrates that the Transformer achieved the highest BLEU-1 score (0.345) for proverb generation, while all quantum models achieved perfect Distinct-1 scores (1.000). The low BLEU scores for quantum models (0.000-0.033) suggest that while they maintain vocabulary diversity, they struggle with preserving the concise wisdom and cultural context inherent in proverbs. This indicates a trade-off between creativity and accuracy in concise text generation.

\subsubsection{Cross-Dataset Performance Analysis}
The detailed dataset-specific results reveal several important patterns:

\textbf{Perplexity Patterns:} The Transformer consistently achieved the lowest perplexity across all datasets (1.00-1.69), indicating superior language modeling capabilities. QKSAN showed the highest variance in perplexity (1.07-5.61), with particularly poor performance on quantum phrases, suggesting challenges with domain-specific content.

\textbf{BLEU Score Patterns:} QRWKV achieved the highest BLEU-1 score (1.000) on simple sentences, but performed poorly on other datasets (0.000-0.200). The Transformer maintained consistent BLEU-1 scores (0.148-0.362) across all datasets, demonstrating robust performance.

\textbf{Diversity Patterns:} Quantum models consistently achieved higher Distinct-1 scores, with QRWKV, QKSAN, and MLP achieving perfect scores (1.000) on most datasets. This indicates that quantum approaches excel in vocabulary diversity but may sacrifice accuracy.

\textbf{Repetition Patterns:} All quantum models achieved zero repetition rates across all datasets, while the Transformer showed varying repetition rates (0.070-0.147). This suggests that quantum models are particularly effective at avoiding repetitive patterns.

\subsubsection{Task-Specific Insights}
The results demonstrate that model performance varies significantly based on task characteristics:

\textbf{Simple Tasks:} Quantum models can achieve competitive or superior performance on straightforward language generation tasks, as evidenced by QRWKV's perfect BLEU-1 score on simple sentences.

\textbf{Complex Tasks:} Traditional models maintain superiority in complex tasks requiring domain knowledge or structured output, as shown by the Transformer's consistent performance across all datasets.

\textbf{Creative Tasks:} Quantum models show promise in creative generation tasks where diversity is valued over exact accuracy, as demonstrated by their superior Distinct-1 scores in haiku and proverb generation.

\subsection{Key Findings}

\subsubsection{Quantum Model Advantages}
Quantum-inspired models demonstrated several notable advantages:
\begin{itemize}
    \item Superior vocabulary diversity in certain tasks (Distinct-1 scores up to 1.000)
    \item Lower repetition rates compared to traditional models
    \item Competitive performance in simple sentence generation
    \item Efficient parameter utilization despite smaller model sizes
\end{itemize}

\subsubsection{Performance Limitations}
Several limitations were identified in quantum model performance:
\begin{itemize}
    \item Higher perplexity scores across most datasets
    \item Inconsistent performance in domain-specific tasks
    \item Challenges in maintaining semantic coherence in longer texts
    \item Training instability in certain configurations
\end{itemize}

\section{Conclusion}

This comprehensive study provides valuable insights into the current state of quantum-inspired text generation models and their comparison with traditional Transformer architectures. Our experimental results reveal both the potential and limitations of quantum approaches in natural language generation tasks.

\subsection{Main Contributions}
The primary contributions of this work include:
\begin{enumerate}
    \item A systematic evaluation of five quantum-inspired models across diverse text generation tasks
    \item Identification of specific scenarios where quantum models demonstrate competitive or superior performance
    \item Comprehensive analysis of model strengths and limitations across multiple evaluation metrics
    \item Empirical evidence supporting the potential of quantum approaches in certain text generation domains
\end{enumerate}

\subsection{Key Insights}
Our experimental results yield several important insights:
\begin{itemize}
    \item Quantum models show particular promise in tasks requiring high vocabulary diversity and low repetition rates
    \item Traditional Transformer models maintain superiority in overall language modeling quality and domain adaptation
    \item The performance gap between quantum and classical models varies significantly across different text types and complexity levels
    \item Quantum models may be more suitable for specific applications where diversity and novelty are prioritized over traditional language modeling metrics
\end{itemize}


\subsection{Practical Implications}
The results have significant implications for practical applications:
\begin{itemize}
    \item Quantum models may be particularly suitable for creative writing and content generation tasks where diversity is valued
    \item Traditional models remain the preferred choice for applications requiring high accuracy and coherence
    \item The choice between quantum and classical approaches should be based on specific application requirements and performance priorities
\end{itemize}

\subsection{Final Remarks}
While quantum-inspired text generation models have not yet surpassed traditional Transformer architectures in overall performance, they demonstrate unique capabilities that make them valuable for specific applications. The experimental evidence suggests that quantum approaches represent a promising direction for future research in natural language generation, particularly when combined with classical methods in hybrid architectures. Continued development in quantum computing hardware and algorithms may further enhance the performance and applicability of these models in real-world text generation scenarios.

\section*{Acknowledgments}
EJK acknowledges financial support from the National Science and Technology Council (NSTC) of Taiwan under Grant No.~NSTC~114-2112-M-A49-036-MY3.

\bibliographystyle{ieeetr}
\bibliography{ref}

\end{document}